\documentclass[sigconf]{acmart}
\usepackage{hyperref}
\usepackage{subcaption}
\usepackage[utf8]{inputenc}
\usepackage{microtype}
\usepackage{bm}
\usepackage{url}
\usepackage{amsthm}
\usepackage[inline]{enumitem}
\usepackage{multirow}
\usepackage{colortbl}
\usepackage{soul}
\usepackage{xspace}

\newcommand{\combinedret}{$f_{\text{Ret}}$\xspace}
\newcommand{\ctrlce}{\textsc{CtrlCE}\xspace} 
\newcommand{\ctrlceit}{\textsc{CtrlCE}$_{\text{It}}$\xspace} 
\newcommand{\ctrlcecv}{\textsc{CtrlCE}$_{\text{CV}}$\xspace} 

\copyrightyear{2025}
\acmYear{2025}
\setcopyright{cc}
\setcctype{by}
\acmConference[SIGIR '25]{Proceedings of the 48th International ACM SIGIR Conference on Research and Development in Information Retrieval}{July 13--18, 2025}{Padua, Italy}
\acmBooktitle{Proceedings of the 48th International ACM SIGIR Conference on Research and Development in Information Retrieval (SIGIR '25), July 13--18, 2025, Padua, Italy}\acmDOI{10.1145/3726302.3729913}
\acmISBN{979-8-4007-1592-1/2025/07}
\acmDOI{10.1145/3726302.3729913}
\ccsdesc[500]{Information systems~Personalization}
\keywords{controllable personalization; cross-encoders; calibrated retrievers}

\begin{document}

\title[Bridging Personalization and Control in Scientific Personalized Search]{Bridging Personalization and Control in\\ Scientific Personalized Search}
\author{Sheshera Mysore}
\affiliation{%
 \institution{University of Massachusetts}
 \country{Amherst, USA}
}
\email{smysore@cs.umass.edu}

\author{Garima Dhanania}
\affiliation{%
 \institution{University of Massachusetts}
 \country{Amherst, USA}
}
\email{garimadhanania@gmail.com}

\author{Kishor Patil}
\affiliation{%
 \institution{Lowe’s Companies, Inc}
 \country{Bangalore, India}
}
\email{kishor.patil@lowes.com}

\author{Surya Kallumadi}
\affiliation{%
 \institution{Coursera}
 \country{Mountain View, USA}
}
\email{surya@ksu.edu}

\author{Andrew McCallum}
\affiliation{%
 \institution{University of Massachusetts}
 \country{Amherst, USA}
}
\email{mccallum@cs.umass.edu}

\author{Hamed Zamani}
\affiliation{%
 \institution{University of Massachusetts}
 \country{Amherst, USA}
}
\email{zamani@cs.umass.edu}

\settopmatter{printacmref=true}

\renewcommand{\shortauthors}{Mysore et al.}
\begin{abstract}
    Personalized search is a problem where models benefit from learning user preferences from per-user historical interaction data. The inferred preferences enable personalized ranking models to improve the relevance of documents to users. However, personalization is also seen as opaque in its use of historical interactions and is not amenable to users' control. Further, personalization limits the diversity of information users are exposed to. While search results may be automatically diversified this does little to address the lack of control over personalization. In response, we introduce a model for personalized search that enables users to control personalized rankings proactively. Our model, \ctrlce, is a novel cross-encoder model augmented with an \emph{editable memory} built from users' historical interactions. The editable memory allows cross-encoders to be personalized efficiently and enables users to control personalized ranking. Next, because all queries do not require personalization, we introduce a \emph{calibrated mixing model} which determines when personalization is necessary. This enables users to control personalization via their editable memory only when necessary. To thoroughly evaluate \ctrlce, we demonstrate its empirical performance in four domains of science, its ability to selectively request user control in a calibration evaluation of the mixing model, and the control provided by its editable memory in a user study.
\end{abstract}

\maketitle

\section{Introduction}
Personalized search powers several industry scale search systems for products \cite{aslanyan2020ecomps, yu2023realetsy}, movies \cite{ostuni2023netflixps}, jobs \cite{hathuc2016linkedinps}, and web-search more broadly \cite{google2009personalization}. While personalization in search systems improves the relevance of search results and increases the uptake of systems, personalized systems are commonly seen as opaque and failing to provide users with sufficient control over personalized predictions \cite{eiband2019peoplealgos, konstan2021human}. Prior work has noted that personalized ranking is more likely to prevent users from seeing the breadth of information in a document collection, raising concerns of fairness \cite{chien2023fairness} and hindering proactive exploration in learning-oriented applications such as education and science \cite{salehi2015acadperson}.

Such concerns about personalization have been addressed through two avenues: diversifying search results and enabling interactive control over personalized ranking. While diversification of search results is meaningful \cite{radlinski2006psdiverse, vallet2012psdiverse}, it does not improve user control or facilitate proactive user-driven interaction and discovery \cite{ruotsalo2018perexplore}. To remedy this, a small body of work has explored providing users interactive control over personalized search by rendering user representations used for personalization ``editable'' \cite{ahn2015openusermodel, zemede2017edps}. However, this work has focused on designing visualization interfaces for user control with simpler token/entity-based user representations. While prior work on controllable personalization for \emph{search} has been limited, a significant amount of work has explored scrutable/controllable approaches for personalized \emph{recommendation}. This work has explored technical approaches for scrutable recommendations \cite{balog2019transparent, konstan2021human} and run human-centered evaluations to show how control over personalized recommendations improved user satisfaction and trust \cite{jin2017different, jannach2016user}. In this paper, we take inspiration from this work and extend the technical body of work on controllable personalized search.

We begin by outlining the following goals for controllable personalized search: (1) to allow control, the user representations used for personalization must be transparent to users and should enable users to express preferences through intuitive profile edits. (2) Since search can be performed without any personalization a controllable model should enable users to opt-out of personalization, supporting ``no personalization'' and (3) Since only some queries are likely to require personalization \cite{ai2019zam, teevan2008pernoper} a controllable model should highlight the queries for which user profile control would be meaningful. 

\begin{figure}[t]
     \centering
     {\includegraphics[width=0.3\textwidth]{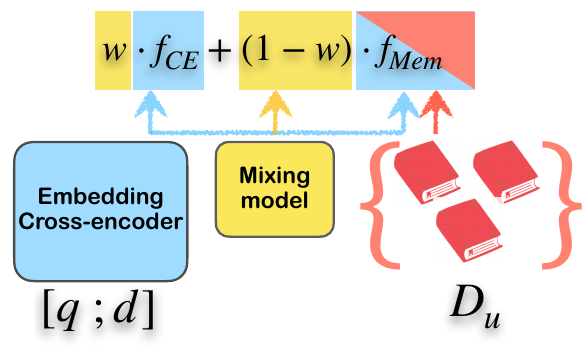}}
     \caption{Our approach \ctrlce, augments a cross-encoder with an editable user profile using a calibrated mixing model. Our training procedure ensures that the mixing models score $w$ remains proportional to the performance of $f_{\text{CE}}$. This ensured that it can be used for seeking edits to a user profile only when necessary.}
     \label{fig-high-level}
     \vspace{-0.5cm}
 \end{figure}

To fulfill these goals we introduce \ctrlce (see Figure \ref{fig-high-level}), a controllable cross-encoder model personalized with an \emph{editable memory} constructed from historical user documents. We explore two multi-vector user memories, an item-level memory and a novel concept-based memory introduced in recent work on controllable recommendation \cite{mysore2023lace}. While both representations remain transparent and editable, concept-based memories offer a richer set of edit operations than item-based memories -- we detail this in \S\ref{sec-edps-memce-inference}. Next, to ensure that the \ctrlce cross-encoder benefits, both, from the rich query-document interaction common in cross-encoders and from interaction with the editable user memory we formulate it as a novel \emph{embedding cross-encoder}. This diverges from standard \texttt{CLS} token-based cross-encoders and instead learns separate but contextualized query and document embeddings. Finally, we train a novel \emph{calibrated mixing model} which intelligently combines query-document and user-document scores while also identifying the queries that are likely to benefit from user profile edits.

In experiments on datasets of personalized search from four scientific domains \ctrlce outperforms standard personalization approaches based on dense retrieval, personalized ensemble models, and non-personalized approaches spanning sparse, dense, and cross-encoder retrievers/re-rankers by 6.4-10.6\% across evaluation metrics. Then we demonstrate that \ctrlce fulfills the goals of controllability: empirically demonstrating its ability to perform with ``no personalization'', showing in calibration evaluations that it effectively identifies queries that need personalization e.g. under-specified and exploratory queries, and showing in a user study that performance can be improved through interaction with the user profile.
To the best of our knowledge \ctrlce is the first approach for controllable personalized search with language model based cross-encoders and extends an under-explored area. We release code and data at: \url{https:/github.com/iesl/controllable-personalization-ctrlce}

\section{Related Work}
\label{sec-edps-relatedwork}

\textbf{Personalized search.} Personalization has broadly been explored in search over personal document collections and referred to as ``personal search'' \cite{bi2021userbehavior, jiang2017photovideo} and search over a shared document collection \cite{taveen2005personalizedsearch}. Work on personal search commonly suffers from under-specified or context-dependent queries and sparse user document interactions. Therefore, prior work has focused on leveraging metadata \cite{bendersky2017attribute} and contextual information such as time and location of queries to improve performance \cite{zamani2017contextualps, qin2020matchcross} or on training schemes to learn from sparse interactions \cite{wang2016positionbias, bi2021userbehavior}. On the other hand, personalized search has focused on constructing user models from users' historical interactions and using them for re-ranking documents -- this is more relevant to our work. 

To build user representations early work leveraged term level models \cite{taveen2005personalizedsearch}, topic models \cite{sontag2012probps}, and latent representations learned through matrix factorization \cite{cai2014bprps}.
More recent work has learned personalized word embeddings \cite{yao2020wordembps}, and learned user representations with RNNs \cite{ai2017heirarchicalrnn, ai2019zam} and transformers \cite{bi2020tfzam}. Current work has leveraged dense retrieval models fine tuned for personalized re-ranking \cite{zhou2021pssl, zeng2023personalizeddense}. In relying on pre-trained language models \ctrlce resembles these approaches. However, prior work leverages attention and shallow transformer layers for query-document and user-document scoring. In contrast, our work leverages a query-document cross-encoder delivering stronger performance -- we show this in Section \ref{sec-edps-expresults}. \citet{dai2023contrastivecross} present an exception and leverage cross-encoders that input \emph{all} historical context, query, and documents for personalized product search. Notably, this is only feasible with short historical texts, queries, and documents common in product search. In Section \ref{sec-edps-noperson} we also show its inability to be controllable. 

\textbf{Controllable personalized search.} While the above approaches explore personalized search, they don't consider control over personalization. The most relevant prior work is provided by \citet{zemede2017edps} and \citet{ahn2015openusermodel} -- both of who explore visualization interfaces for interacting with term/entity-based user profiles in personalized search. While meaningful, this work does not explore performant retrieval models as ours does. Finally, recent work \cite{wang2024pnd, liu2024divpers} explores combining personalization and search result diversification by learning when personalization is necessary while sharing in our motivation these approaches do not enable interactive control by users through user profile edits as we do. 

In this regard, \citet[LACE]{mysore2023lace} who introduce editable user profiles for recommendation tasks presents closely related work to \ctrlce. However, our work differs substantially: we focus on search tasks which bring additional challenges for controllability compared to recommendation setups that lack user queries (Section \ref{sec-edps-problem}), \ctrlce introduces a calibrated mixing model which highlights queries where control is necessary, and we show how performant cross-encoder models can be controllably personalized. Additionally, our ablation experiments compare to LACE in Section \ref{sec-edps-ablations} and show significantly improved performance by \ctrlce.

\textbf{Calibrated retrievers.}
The calibrated mixing model used in \ctrlce ties our work to a small body of work on calibrated ranking models -- this work aims to train ranking models that produce confidence scores alongside relevance scores or produces scores that are proportional to ranking model performance, our work relates to the latter. While we build on \citet{le2022scalecalib} who train scale-calibrated ranking models for CTR systems, we leverage scale-calibration for facilitating user control in personalized search. Other work has explored probabilistic uncertainty estimation in retrievers through joint training of retrievers in RAG systems \cite{dhuliawala2022mrcalibration}, Monte Carlo dropout \cite{cohen2021uncertainrel}, and Gaussian query and document embeddings \cite{zamani2023multivar}. In contrast with probabilistic uncertainty estimation, our mixing model produces calibrated scores through regularization and does not require extensive changes to training, model architecture, or additional inference costs. Finally, our mixing model may be seen as a query performance prediction (QPP) model \cite{arabzadeh2024qpptut} -- in this sense, our work represents the first approach using a QPP model for improving personalized search.

\section{Problem Formulation}
\label{sec-edps-problem}
We consider a personalized search problem where a user $u \in \mathcal{U}$,  submits a query $q$ to a retrieval system, for retrieving documents from a document collection $\mathcal{D}$ shared with other users. Each user is associated with a user profile $\mathcal{P}_u$ constructed from their historical documents $D_u = \{d_u^{i}\}_{i=1}^{N_u}$. Historical documents are assumed to capture users' interests and may be documents that users have authored, read, clicked, etc in the past. The retrieval system, \combinedret, is tasked with producing a ranking over $\mathcal{D}$ personalized to the user $u$ as $R_u = f_{\text{Ret}}(\mathcal{P}_u, q, \mathcal{D})$. In practice, we take \combinedret to be a re-ranking model ranking top $K$ documents from a first-stage retriever.

Given the value of control over personalization \cite{jin2017different, jannach2016user, sciascio2020ctrltransp}, we are interested in allowing users control over $R_u$ by manipulating $\mathcal{P}_u$. Such a controllable model should fulfill the following goals:

\ul{D1: Communicate interests to the user}: The profile should be readable by users to allow edits to it. Specifically, the profile $\mathcal{P}_u$ must communicate the interests of $u$ as represented in $D_u$. Importantly, we only require our transparent user profile to facilitate interactive control over $R_u$, without requiring a fully transparent model.

\ul{D2: Control retrieval via profile interactions}: The profile and model should support edit operations which are reflected in the rankings over documents $R'_u = f(\mathcal{P}'_u, q, \mathcal{D})$. While $R_u$ may trivially be updated via edits to the query, profile interactions can change longer term interests \cite{bennett2012longshort}, or allow clarifications in complex exploratory searches \cite{ruotsalo2018perexplore} where changes to the query may not be obvious. Finally, we require \combinedret to support retrieval in the absence of any personalization $f_{\text{Ret}}(q, \mathcal{D})$ given that some users might not desire personalization for search or have any historical documents. 

\ul{D3: Solicit user input when necessary}: We require \combinedret to identify when profile interactions are likely to be meaningful so that user feedback can be obtained only when necessary. This follows from findings that all queries do not require personalization \cite{ai2019zam, teevan2008pernoper}. Therefore, user edits to $\mathcal{P}_u$ may not always be beneficial.

\ul{D4: Performant retrieval}: Finally, we require performant retrieval before and after profile interactions since users desire a balance between automated predictions and control \cite{knijnenburg2011energy, Yang2019intention}.

\textbf{Profile Design.} For $\mathcal{P}_u$, we explore two designs -- (1) A concept based user profile consisting of a set of natural language concepts, $\mathcal{P}_u=\{ k_1, \dots k_P\}$ inferred from $D_u$, and (2) An item based user profile, $\mathcal{P}_u=\{ d^1_u, \dots d^P_u\}$, which directly represents user interests with $D_u$. Compared to item based profiles, concept based profiles are more succinct and readable -- The concepts represent sets of items enabling efficient user interaction and provide short natural language descriptions improving readability. They have also been found to be an intuitive representation for user interaction in prior work \cite{chang2015setprefs, balog2019transparent}. On the other hand, item based representations are finer grained and promise stronger performance. However, both designs are controllable and allow users to express their preferences for controlling personalization -- we discuss this in Section \ref{sec-edps-memce-inference}.

\section{Proposed Approach}
\label{sec-edps-proposed}
For controllable personalized search, we present \ctrlce, a language model based cross-encoder model personalized using an editable memory constructed from user items. We base \ctrlce on cross-encoders because of their strong performance in search tasks, strong generalization ability across domains \cite{thakur2021beir}, and standard use as re-ranking models \cite{lin2021pretrained}. To personalize our cross-encoder, we augment it with a \emph{editable memory} constructed from a user's historical documents. To ensure that \ctrlce remains controllable and performant we introduce three key novelties: (1) We introduce an \textbf{embedding cross-encoder} which learns separate yet \emph{contextualized} query and document embeddings allowing the \ctrlce cross-encoder to interact with a multi-vector editable user memory. This is in contrast to standard cross-encoders which learn a \emph{fused} query-document representation from \texttt{CLS} tokens of pre-trained language models. 
(2) We construct \textbf{editable user profiles} based on dense retriever embeddings. We consider two user profile designs, concept and item-based (see Section \ref{sec-edps-problem}). While item-based user representations are naturally transparent and editable by users, we introduce \emph{concept-value user representations} for our concept-based user profiles. This user representation pairs each concept in a user profile with a \emph{personalized concept value} computed from user documents. The concept values may be seen as labeled cluster centroids of user documents. 
(3) Finally, since we are interested in obtaining user edits to $\mathcal{P}_u$ only when necessary, we introduce a \textbf{calibrated mixing model} which learns to combine query-document scores from a cross-encoder with user-document scores obtained from user profiles (Figure \ref{fig-high-level}). Besides combining these scores, the mixing models' scores highlight the queries that are likely to benefit from personalization and may in turn benefit from user edits. Next, we describe \ctrlce, the editable memory/user profile, and the training procedure that results in our calibrated mixing model.

\subsection{Memory Augmented Cross-encoder}
\label{sec-edps-memcrossenc}
\subsubsection{Model Overview}
\label{sec-edps-memcrossenc-overview}
We formulate \ctrlce as a re-ranking model for documents: $s_d = f_{\text{Ret}}(\mathcal{P}_u, q, d)$. We assume access to a user query $q$, a user profile $\mathcal{P}_u$ constructed from user documents $D_u$, and the top $K$ candidate documents $d\in\mathcal{D}$ from a first stage ranking model. \ctrlce computes document relevance ($s_d$) using two scores, query-document relevance from an embedding cross-encoder $f_{\text{CE}}$ and user-document relevance from user memory $f_{\text{Mem}}$. These are combined with a calibrated mixing model $g_{\text{Mix}}$:
\begin{equation}
    s_d = w\cdot s_d^q + (1-w)\cdot s_d^u = w\cdot f_{\text{CE}}(q,d) + (1-w)\cdot f_{\text{Mem}}(\mathcal{P}_u, q, d)
    \label{eq-edps-highlevelscore}
\end{equation}
Here, $f_{\text{CE}}$ is formulated as an embedding cross-encoder \cite{yadav2022anncur} and $f_{\text{Mem}}$ computes a score for candidate $d$ based on the interaction between a cross-encoded document embedding and the user profile/memory. Note that in contrast with standard cross-encoders \cite{lin2021pretrained}, an embedding cross-encoder learns \emph{separate} yet cross-encoded query and document representations, $\mathbf{q}$ and $\mathbf{d}$. This allows $f_{\text{CE}}$ to interact with $\mathcal{P}_u$ in $f_{\text{Mem}}$. Therefore, \ctrlce benefits from the strong performance of cross-encoders and yet allows a cross-encoder to interact with user memory for personalization. We formulate $f_{\text{Mem}}$ as a multi-vector user representation, i.e.\ $\mathcal{P}_u$ is represented as a set of $P$ vectors $\mathbf{V}_{i=\{1\dots P\}}$.

Next, the two scores are combined using weight $w$ from a mixing model $g_{\text{Mix}}$. This is trained with a calibrated training objective (Section \ref{sec-edps-memce-training}) which ensures that $w$ remains proportional to the ranking performance of $f_{\text{CE}}$ and serves as a \emph{performance predictor} for it. This allows $g_{\text{Mix}}$ to be used by system designers to obtain user edits when \ctrlce relies more on $f_{\text{Mem}}$ (D3 of Section \ref{sec-edps-problem}). Finally, decomposing $s_d$ into two scores supports ``no personalization'' by dropping the user-document score, $s_d^u$ (D2 of Section \ref{sec-edps-problem}).

\begin{figure}[t]
     \centering
     {\includegraphics[width=0.4\textwidth]{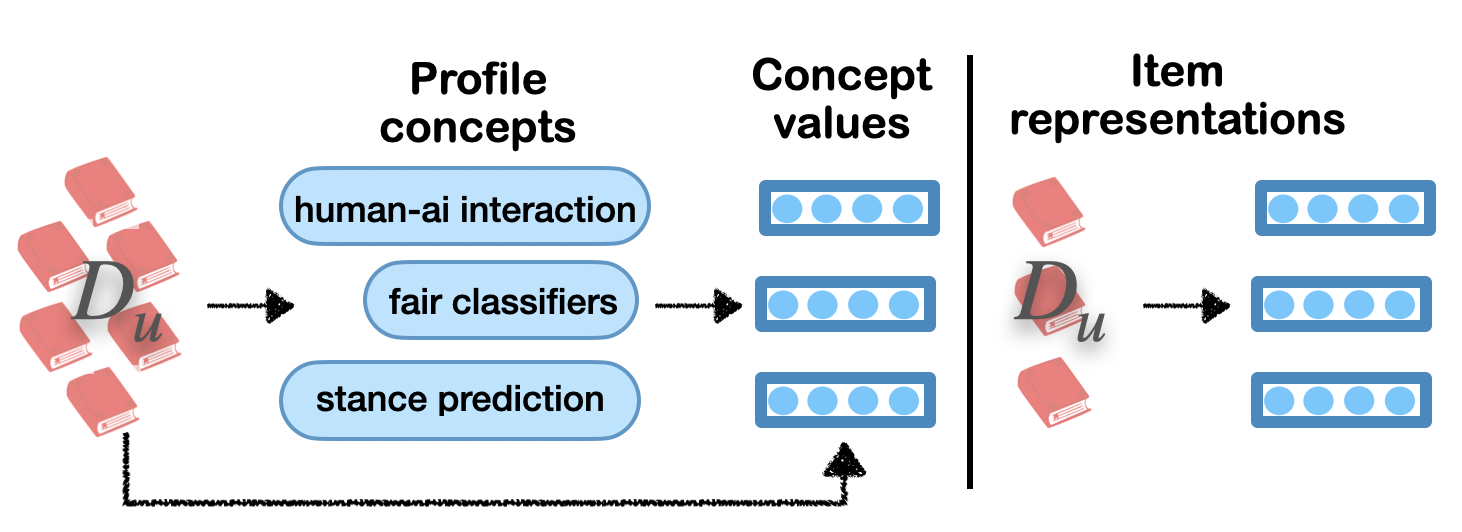}}
     \caption{Concept-value memories represent users with concepts and their personalized concept values. Item memories directly represent users with item representations.}
     \label{fig-user-memory}
     \vspace{-0.5cm}
 \end{figure}

Specifically, $f_{\text{CE}}$ inputs concated query and document text into a pre-trained language model (LM) encoder $\texttt{Enc}_{\text{CE}}([q;d])$ and computes the dot product of cross-encoded query and document embeddings: $f_{\text{CE}}(q,d) = \mathbf{q}^T\mathbf{d}$. Both $\mathbf{q}$ and $\mathbf{d}$ are computed as contextualized token embedding averages. Next, user memory embeddings $\mathbf{V}_{i=\{1\dots P\}}$ are computed from user documents $D_u$ using a pre-trained LM-based memory encoder $\texttt{Enc}_{\text{Mem}}$ in an offline memory construction stage (see Section \ref{sec-edps-cvmem},\ref{sec-edps-itmem}). Then, $f_{\text{Mem}}$ is computed as: $\texttt{max}_{i\in\{1\dots P\}} \mathbf{d}^T\mathbf{V}_i$. Finally, the mixing weights $w$, are obtained as a function of the query $\mathbf{q}$, the length of the user profile and query (in tokens), and the model scores $s_d^q$ and $s_d^u$ as: $w=\texttt{sigmoid}(\text{MLP}([\mathbf{q}, \texttt{len}(q), P, s_d^q, s_d^u]))$. This design builds on the intuition that these features are likely to help predict the performance for $f_{\text{CE}}$ building on prior work in query performance prediction for non-personalized search \cite{arabzadeh2024qpptut}. This gives:
\begin{align}
    s_d &= w\cdot\mathbf{q}^T\mathbf{d} + (1-w)\cdot\texttt{max}_{i\in\{1\dots P\}} \mathbf{d}^T\mathbf{V}_i \label{eq-edps-detailedscore}\\
     w &= g_{\text{Mix}}(\mathbf{q}, \texttt{len}(q), P, s_d^q, s_d^u)\label{eq-edps-mixing}
\end{align} 
Next, to ensure that the user profile remains transparent and editable by users (D1, D2) we leverage either a concept-value user representation or an item-based user representation (see Figure \ref{fig-user-memory}).

\subsubsection{Concept Value Memories.}
\label{sec-edps-cvmem}
Our concept-value memories represent users with a concept-based user profile $\mathcal{P}_u$ containing $P$ natural language concepts: $\mathcal{P}_u=\{ k_1, \dots k_P\}$ -- the concepts are succinct descriptors for user documents and are inferred for $D_u$ using dense retrieval of concepts for each user document from a large inventory of interpretable concepts (e.g.\ Wikipedia categories). Next, the concepts in $\mathcal{P}_u$ are paired with \emph{personalized} concept-values $\mathbf{V}_{i=\{1\dots P\}}$ which are computed as a function of $D_u$ and $\mathcal{P}_u$. While concepts may be represented simply as embeddings of the concept text, personalized concept values enable stronger personalization performance by using the user documents to compute concept embeddings. Further, since the personalized concept value is also a function of the concept text, any user edits to the concept text also update the personalized concept value. To construct the concept-value memories we use Optimal Transport (OT), a linear programming method for computing assignments between sets of vectors \cite{peyre2019computational}, to make a sparse assignment of user documents to concepts. The assigned document content is then used to compute the personalized concept values (Equation \eqref{eq-concept-value-computation}). Our work builds on prior work \cite{mysore2023lace} which introduces concept-value memories for controllable recommendations. Here, we show how they can enable control over personalized search models based on powerful cross-encoders.

Specifically, to construct the concept-value memories, we begin by assuming access to a large concept inventory $\mathcal{K}$ (detailed in Section \ref{sec-edps-expsetup}), and an encoder ($\texttt{Enc}_{\text{Mem}}$) for concepts and user documents $D_u$ that outputs embeddings for them as $\mathbf{K}$ and $\mathbf{S}_D$. Then, $\mathcal{P}_u$ is constructed by retrieving the top-$P$ concepts from $\mathcal{K}$ for the documents in $D_u$ based on their embeddings -- a form of zero-shot classification. Notably, $P < N_u$ ensures that the concept-based profile represents a succinct and readable representation of user documents. Next, a sparse and soft assignment of the documents to concepts $\mathbf{Q}_{D\rightarrow\mathcal{P}}$ is computed using optimal transport which solves the linear assignment problem: $argmin_{\mathbf{Q}'} \langle dist(\mathbf{S}_D, \mathbf{K}_{\mathcal{P}})\cdot \mathbf{Q}'\rangle$. In the interest of space, we refer readers to prior work for a more detailed presentation of OT \cite{mysore2023lace, peyre2019computational}. Next, $\mathbf{V}_i$ is computed as an assignment-weighted average of the document content:
\begin{align}
    \mathbf{V}_{i=\{1\dots P\}} = \frac{1}{\sum_{j=1}^{N_u}\mathbf{Q}_{ji}}\sum_{j=1}^{N_u}\mathbf{Q}_{ji}\cdot\mathbf{S}_j
    \label{eq-concept-value-computation}
\end{align}

The above design presents some important benefits for controllability, OT computes sparse assignments $\mathbf{Q}$ \cite{peyre2019computational}, ensuring that every user document is only assigned to a small number of relevant concepts. Therefore, the concepts partition the documents into soft clusters ``tagged'' by their concept. This enables users to specify positive or negative preferences for specific concepts, which includes or excludes clusters of documents in generating personalized rankings. Further, user edits to the text of concepts influences their embeddings $\mathbf{K}_{\mathcal{P}}$, which in turn influence $\mathbf{Q}_{D\rightarrow\mathcal{P}}$, $\mathbf{V}_{i=\{1\dots P\}}$, and $R_u$ - allowing edits to reflect in rankings. Finally, OT is readily solved using the Sinkhorn algorithm \cite{cuturi2013sinkhorn} which runs efficiently on GPUs and can be used inside models trained with gradient descent.

\subsubsection{Item Memories.}
\label{sec-edps-itmem}
In contrast with concept-value memories, item memories are a simpler user representation where items in $D_u$ are used directly as a user representation, resulting in $\mathbf{V}_{i=\{1\dots P\}} = \mathbf{S}_D$ with $P = N_u$. Here $\texttt{Enc}_{\text{Mem}}$ directly yields embeddings for $D_u$. Item memories allow users only to control $s_d^u$ through including or excluding items in $\mathcal{P}_u$. While this can result in cumbersome edits for large profiles, item memories retain finer grained item representations compared to the aggregated representations of concept-value memories which are likely to offer better performance in search tasks. Our experiments (Section \ref{sec-edps-expresults}) demonstrate the efficacy of both item and concept-based memories.

\subsection{Training}
\label{sec-edps-memce-training}
We propose a two-stage procedure for training the embedding cross-encoder ($\texttt{Enc}_{\text{CE}}$) and the calibrated mixing model $g_{\text{Mix}}$. In stage-1 we train our embedding cross-encoder $\texttt{Enc}_{\text{CE}}$ while omitting the mixing model. This results in the relevance scores: $s_d = s_d^q + s_d^u$. Then, in stage-2, we introduce mixing model $g_{\text{Mix}}$ to combine $s_d^q$ and $s_d^u$ per Equation \eqref{eq-edps-highlevelscore}: $s_d = w\cdot s_d^q + (1-w)\cdot s_d^u$. We use historical user interaction data and a pairwise cross-entropy loss to train $\texttt{Enc}_{\text{CE}}$ and a scale calibrating cross-entropy loss to train $g_{\text{Mix}}$. Our calibrating training procedure ensures that the scores from $g_{\text{Mix}}$ don't lie at the extremes of its score range, a known property of MLP-based scoring functions \cite{guo2017calibnn, yadav2022anncur, menon22defensedual}, and instead closely tracks the performance of $f_{\text{CE}}$.

Specifically in stage-1, for a user $u$ and query $q$, we assume access to a relevant document $d^+$ and a set of $M$ irrelevant documents $d^-$. This results in a vector of predicted scores $\mathbf{s}_q=[s_{d^+}\dots s_{d^-}]$ and binary relevance labels $\mathbf{y}_q=[1\dots 0]$. To train $\texttt{Enc}_{\text{CE}}$ we minimize the standard softmax (\texttt{sm}) cross-entropy loss:
\begin{align}
    \mathcal{L}(\mathbf{y}_q, \mathbf{s}_q) = -\sum_{i=1}^{M}\mathbf{y}_{q}[i]~\text{log}~\texttt{sm}(\textbf{s}_{q}[i])
\end{align}
The parameters of $\texttt{Enc}_{\text{CE}}$ are updated while memory representations and $\texttt{Enc}_{\text{Mem}}$ are kept fixed throughout training to ensure that model training remains scalable on our GPUs. 
We initialize $\texttt{Enc}_{\text{Mem}}$ with a strong pre-trained LM encoder optimized for dense retrieval and $\texttt{Enc}_{\text{CE}}$ with a pre-trained LM encoder.

In stage-2 we train $g_{\text{Mix}}$. Here, we freeze $\texttt{Enc}_{\text{Mem}}$ and $\texttt{Enc}_{\text{CE}}$ and obtain personalized ranking scores per Equation \eqref{eq-edps-highlevelscore}.  While we use data identical to that used for training $\texttt{Enc}_{\text{CE}}$, we modify the softmax objective and instead leverage a scale calibrating softmax objective \cite{le2022scalecalib}. This modifies the softmax loss by adding an ``anchor'' example with target score $y_0 \in [0,1]$, which is a tunable hyperparameter, and logit $s_0$ set to $0$. This results in $\mathbf{s}_q'= [s_{d^+}\dots s_{d^-}, s_0]$ and $\mathbf{y}_q' = [1-y_0 \dots 0, y_0]$ and the loss:
\begin{align}
    \mathcal{L}(\mathbf{y}_q', \mathbf{s}_q') = - \sum_{i=1}^{M} \mathbf{y}_{q}'[i]~\text{log}\frac{e^{\mathbf{s}_{q}'[i]}}{\sum_{j}e^{\mathbf{s}_{q}'[j]}+1} + y_0~\text{log}~({\sum_{j}e^{\mathbf{s}_{q}'[j]}+1})
    \label{eq-calibrated-expansion}
\end{align}
The insertion of the anchor target $y_0$ regularizes the scores $s_d$ -- penalizing large scores (second term) and preventing smaller scores from being lowered (first term) -- ensuring that scores are driven away from the extremes of the score distribution. Further, since only $g_{\text{Mix}}$ is trained with the calibrated objective the weights $w$ more smoothly tradeoff the query-document scores $s_d^q$ and the user-document scores $s_d^u$. As we show in Section \ref{sec-edps-calibeval}, this results in $w$ being better calibrated with the performance of $s_d^q$. This calibrated weight $w$ may be used to guide users toward making profile interactions when $w$ indicates that $s_d^u = f_{\text{CE}}(q,d)$ will perform poorly.

\subsection{Inference}
\label{sec-edps-memce-inference}
\textbf{Retrieval.} Performing retrieval with \ctrlce follows a standard two-stage ranking procedure, a first stage ranker retrieves a set of $K$ documents from $\mathcal{D}$, then \ctrlce functions as a re-ranker. It uses $s_d$ (Equation \eqref{eq-edps-detailedscore}) to re-rank the top $K$ documents and produce a personalized ranked list $R_u$. To ensure that \ctrlce can be run on standard GPUs, \ctrlce is implemented using 110M parameter transformer LMs, and $g_{\text{Mix}}$ is implemented as a 1-layer MLP. Further, because we formulate $f_{\text{CE}}$ as an embedding cross-encoder (Section \ref{sec-edps-memcrossenc-overview}), \ctrlce may be scaled readily to the scale of embedding-based dense retrieval models using recently introduced matrix factorization techniques \cite{yadav2022anncur} -- we leave this to future work.

\textbf{Interactive control.} Control over personalization in \ctrlce is achieved by interactions with the concept-value or item-based user profile.
Given that only some queries may benefit from personalization \cite{ai2019zam, teevan2008pernoper} system designers may only solicit user edits to $\mathcal{P}_u$ for low values of $w$  from the calibrated mixing model $g_{\text{Mix}}$. For example, highly specific lookup queries may not require personalization, on the other hand, exploratory or under-specified queries commonly benefit from personalization \cite{dumais2003refinding}. In addition to these interactions, users may also choose to have ``no personalization''. In \ctrlce, this may be accomplished by re-ranking documents based on query-document relevance ($s_d^q$) alone.

In interacting with $\mathcal{P}_u$, item profiles offer a more limited range of interactions than concept-value profiles. Item memories support positive/negative selections, whereas concept-value memories support both positive/negative selections and profile edits. \ul{1. Positive/negative selection.} Users may choose to include or exclude concept-values (or item embeddings), $\mathbf{V}_{i=\{1\dots P\}}$ ($\mathbf{V}$ in short), to be used for computation of $R_u$. Positive selection results in the positively selected values, $\mathbf{V}[p,:]$ being used for computing $R_u$. Similarly, negative selection results in a compliment of the selections $\mathbf{V}[\overline{n},:]$ being used for computing $R_u$. Such interactions allow users to include/exclude sets of items or individual items from being used to compute user-document scores $s_d^u$. \ul{2. Profile edits.} Further, for concept-value memories users may also directly change the text of concepts in $\mathcal{P}_u$ triggering re-computation of $\mathbf{V}$ i.e. a reorganization of documents in $D_u$. These edits may be to edit incorrectly inferred concepts in $\mathcal{P}_u$ or add missing concepts. In experiments, we refer to the item and concept-value profile variants of \ctrlce as \ctrlceit and \ctrlcecv respectively. 

Finally, note that the design of \ctrlce allows efficient updates to $R_u$ based on both ``no personalization'' and the profile interactions outlined above. This follows from all the score computations in \ctrlce being based on dot-products (Equation \eqref{eq-edps-detailedscore}). Further, our use of an embedding cross-encoder, and our multi-vector user profiles means that representations can be cached per query and updated rankings produced only using efficient computations.

\section{Experiments}
\label{sec-edps-results}
We evaluate \ctrlce on four datasets of personalized search constructed from four different scientific domains in a public benchmark for personalized search in Section \ref{sec-edps-expresults}. Section \ref{sec-edps-ablations} presents a series of ablations for the components in \ctrlce, and Section \ref{sec-edps-inteval} we evaluates the controllable components of \ctrlce.

\subsection{Experimental Setup}
\label{sec-edps-expsetup}

\subsubsection{Datasets} 
\label{sec-exp-datasets}
We use a public benchmark for personalized search \cite{bassani2022person}, consisting of queries, user documents, and document collections from four scientific domains: computer science (Comp Sci), physics (Physics), political science (Pol Sci), and psychology (Psych). Train and test splits are created temporally such that the test set consists of the most recent queries across the dataset. Each dataset contains 150k-500k training queries, 5k-12k test queries, and profiles between 20-300 documents. We use the title and abstract text to represent documents and report performance using NDCG@10 and note that MRR follows identical trends.

\subsubsection{Baselines}
\label{sec-exp-baselines}
We consider a range of standard personalized and non-personalized baselines spanning sparse retrieval, dense retrieval, cross-encoders, and ensemble methods. Our non-personalized approaches span: sparse retrieval with \ul{BM25}, weakly supervised dense retrieval with \ul{Contriver}, supervised dense retriever trained on 1 billion pairs (\ul{MPNet-1B}, HF: all-mpnet-base-v2), and supervised dense retrieval trained on 250M community question answer sites (\ul{MPNet-CQA}, HF: multi-qa-mpnet-base-cos-v1) -- this is noted to be valuable training data for dense retrievers \cite{menon22defensedual}. Finally, \ul{CrossEnc} is a standard cross-encoder trained on the same data as \ctrlce while ignoring the user documents $D_u$. CrossEnc is initialized with MPNet-base \cite{song2020mpnet}, and the query-document score is produced by passing the \texttt{CLS} representation through an MLP. CrossEnc is the closest comparator to \ctrlce since it is a standard non-personalized cross-encoder. We include the details of baselines in our code repository.
\begin{table}[t]
\centering
\caption{\ctrlce is compared against non-personalized (first block) and personalized (second block) approaches. Bold indicates \ctrlce improvement over CrossEnc and $^{*}$ indicates statistical significance with a two-sided t-test at $p < 0.05$.}
\scalebox{0.81}{
\begin{tabular}{rcccc}
\toprule
 & \multicolumn{1}{c}{Comp Sci} & \multicolumn{1}{c}{Physics} & \multicolumn{1}{c}{Pol Sci} & \multicolumn{1}{c}{Psych}\\ 
\cmidrule(lr){2-2}\cmidrule(lr){3-3}\cmidrule(lr){4-4}\cmidrule(lr){5-5}
Model & \small{NDCG@10} & \small{NDCG@10} & \small{NDCG@10} & \small{NDCG@10}\\
\midrule
BM25 & 0.2245 & 0.2688 & 0.2407 & 0.2393\\
Contriver & 0.1658 & 0.1795 & 0.1932 & 0.1887\\
MPNet-1B & 0.2168 & 0.1885 & 0.2306 & 0.2407\\
MPNet-CQA & 0.1969 & 0.2093 & 0.2153 & 0.2187\\
CrossEnc & 0.2934 & 0.3330 & 0.3102 & 0.3346\\
\midrule
\texttt{rf}(BM25,QA) & 0.2849 & 0.3272 & 0.2894 & 0.3112\\
\texttt{rf}(BM25,MPNet-1B,QA) & 0.3115 & 0.3481 & 0.3141 & 0.3485\\
RetAugCE & \textit{0.3244} & \textit{0.3691} & \textit{0.3384} &  \textit{0.3778}\\
\midrule
\ctrlceit & \textbf{0.3223}$^{*}$ & \textbf{0.3657}$^{*}$ & \textbf{0.3378}$^{*}$ & \textbf{0.3704}$^{*}$\\
\ctrlcecv & \textbf{0.3118}$^{*}$ & \textbf{0.3583}$^{*}$ & \textbf{0.3310}$^{*}$ & \textbf{0.3614}$^{*}$\\
\bottomrule
\end{tabular}
}
\label{tab-edps-person-main}
\end{table}

Our personalized approaches use prior personalized dense retrieval models and personalized cross-encoder models: \ul{\texttt{rf}(BM25, QA)}: A rank-fusion approach that learns a weighted combination of BM25 scores and scores from a Query Attention (QA) based user modeling approach. Importantly, QA is a key component of effective personalization in prior work on personalized search, \cite[HRNN-QA]{ge2018rnnqa}, \cite[ZAM]{ai2019zam}, and \cite[EDAM]{jiang2020qagoogle}. QA scores candidate documents based on their dot product similarity to the weighted average of user documents. The weights for user documents are computed as dot-product attentions between query and document representations from MPNet-1B. \ul{\texttt{rf}(BM25, QA, MPNet-1B)}: This adds dense retrieval scores from MPNet-1B to \texttt{rf}(BM25, QA). \ul{RetAugCE}: A cross-encoder personalized with retrieval-augmentation \cite{salemi2023lamp}. It inputs, query, candidate, and the top-1 document most similar to the query from $D_u$. This approach follows the state-of-the-art personalized cross-encoder model for product search \cite[CoPPS]{dai2023contrastivecross}, however given the longer length of documents in our datasets compared to e-commerce products, we use retrieval to reduce input sequence lengths. We use the MPNet-CQA model for retrieving the top-1 document. Finally, note that we primarily aim to demonstrate that \ctrlce results in strong performance compared to several reasonable and strong baselines 
while remaining controllable (Section \ref{sec-edps-inteval}). We leave the exploration of strategies such as contrastive self-supervised training \cite{dai2023contrastivecross, zhou2021pssl} for establishing SOTA performance in controllable cross-encoder models to future work.

\subsubsection{Implementation Details.}
\label{sec-exp-implementation}
In \ctrlce we initialize $\texttt{Enc}_{\text{CE}}$ with MPNet-base \cite{song2020mpnet} and $\texttt{Enc}_{\text{Mem}}$ with MPNet-CQA for both \ctrlceit and \ctrlcecv. We formulate $g_{\text{Mix}}$ as an MLP with one hidden layer of 386 dimensions and use a $\texttt{tanh}$ non-linearity. For first-stage ranking, we use BM25 and re-rank $K=200$ documents per query. Further, in \ctrlcecv for constructing concept-based user profiles for a concept inventory $\mathcal{K}$ we use a collection of computer science concepts from \cite{lo2020s2orc} and Wikipedia categories for Physics, Pol Sci, and Psych. Our code repository includes additional details.

\subsection{Experimental Results}
\label{sec-edps-expresults}
\textbf{Baseline performance.} We begin by examining baseline performance in Table \ref{tab-edps-person-main}. We see that personalized models that ensemble various sparse and dense personalized and non-personalized models with rank fusion (\texttt{rf($\cdot$)}) outperform non-personalized sparse and dense models (rows BM25 to MPNet-CQA). However, a non-personalized cross-encoder (CrossEnc) outperforms all other non-personalized models and approaches the performance of personalized models based on Query Attention (QA). The strong performance of \emph{non-personalized} cross-encoders in personalized search has also been noted in prior work \cite{macavaney2022aolia}. Next, we note that a cross-encoder personalized with retrieval augmentation, RetAugCE outperforms the non-personalized CrossEnc while incurring greater inference costs.

\textbf{\ctrlce performance.} For the proposed \ctrlce models we first note that both \ctrlce variants outperform the non-personalized CrossEnc with improvements of 6.4-10.6\% across evaluation metrics and datasets. This indicates the proposed memory-augmented cross-encoder to be effective at personalization. Next, we compare \ctrlce models to personalized methods based on Query Attention (QA). Here, we note \ctrlce models to outperform personalized ensemble methods. Finally, while \ctrlce performs at par with RetAugCE (no statistically significant difference with \ctrlceit), it does not outperform it. However, as we show in Section \ref{sec-edps-inteval}, RetAugCE does not allow control over personalization -- lacking the ability to identify when control interactions are necessary and being unable to support the ``no personalization'' action (Sections \ref{sec-edps-calibeval} and \ref{sec-edps-noperson}). Therefore, \ctrlce performs at par with state-of-the-art approaches while remaining controllable.

\textbf{\ctrlceit vs \ctrlcecv.} Here, we examine the difference between item and concept-value memories, \ctrlcecv and \ctrlceit. We see \ctrlceit to consistently outperform \ctrlceit by a small margin. 
We hypothesize that this is due to the nature of the search task where most queries seek specific items rather than being exploratory. As a consequence, the finer-grained item representations that \ctrlceit retains allow higher precision in retrieval at the expense of more tedious interactions for controllable personalization. However, recall from Sections \ref{sec-edps-problem} and \ref{sec-edps-memce-inference} that the concept-based profiles of \ctrlcecv provide a richer set of profile interactions and a more compact and readable user profile -- in applications where this is important practitioners may choose to use \ctrlcecv over \ctrlceit. Next, we illustrate the performance resulting from the various components of \ctrlce in a series of ablations.

\subsection{Ablation Study}
\label{sec-edps-ablations}
Table \ref{tab-edps-person-ablation} presents an ablation indicating the performance of the various model components of \ctrlce. We present results for both \ctrlceit and \ctrlcecv. Further, we only present results with NDCG@10 in the interest of space, noting that MRR follows the same trends. We report statistical significance compared against \ctrlceit/\ctrlcecv with $^*$ using a two-sided t-test at $p < 0.05$.

\begin{table}[t]
\centering 
\caption{\ctrlce components ablated for item (\ctrlceit) and concept-value (\ctrlcecv) memories.}
\scalebox{0.85}{
\begin{tabular}{lcccc}
\toprule
 & \multicolumn{1}{c}{Comp Sci} & \multicolumn{1}{c}{Physics} & \multicolumn{1}{c}{Pol Sci} & \multicolumn{1}{c}{Psych}\\ 
\cmidrule(lr){2-2}\cmidrule(lr){3-3}\cmidrule(lr){4-4}\cmidrule(lr){5-5}
Model & \small{NDCG@10} & \small{NDCG@10} & \small{NDCG@10} & \small{NDCG@10}\\
\toprule
\ctrlceit & {0.3223} & {0.3657} & {0.3378} & {0.3704}\\
-- no $f_{\text{Mem}}$ & 0.3010$^*$ & 0.3440$^*$ & 0.3164$^*$ & 0.3465$^*$\\
-- no $f_{\text{CE}}$ & 0.1994$^*$ & 0.2424$^*$ & 0.1997$^*$ & 0.2450$^*$\\ 
-- no $g_{\text{Mix}}$ & 0.3078$^*$ & 0.3452$^*$ & 0.3236$^*$ & 0.3517$^*$\\
-- no calibration & 0.3065$^*$ & 0.3633 & 0.3277$^*$ & 0.3623$^*$\\
\midrule
\ctrlcecv & {0.3118} & {0.3583} & {0.3310} & {0.3614}\\
-- no $f_{\text{Mem}}$ & 0.2936$^*$ & 0.3344$^*$ & 0.3151$^*$ & 0.3368$^*$\\
-- no $f_{\text{CE}}$ & 0.1675$^*$ & 0.1981$^*$ & 0.1792$^*$ & 0.2060$^*$\\
-- no $g_{\text{Mix}}$ & 0.3061 & 0.3390$^*$ & 0.3278 & 0.3399$^*$\\
-- no calibration & 0.3121 & 0.3602 & 0.3326 & 0.3625\\
\bottomrule
\end{tabular}
}
\label{tab-edps-person-ablation}
\end{table}
\textbf{No user-document score.} We begin by examining a test time only change -- after training \ctrlce as described in Section \ref{sec-edps-memce-training}, we produce personalized rankings $R_u$ only using the query-document scores produced by $f_{\text{CE}}$ per Equation \eqref{eq-edps-highlevelscore} (-- no $f_{\text{Mem}}$, Table \ref{tab-edps-person-ablation}). We see that both \ctrlceit and \ctrlcecv see consistent drops in performance indicating the value provided by personalization with editable user memory. 

\textbf{No query-document score.} Having trained \ctrlce, we examine the personalized rankings produced using only the user-document scores produced by $f_{\text{Mem}}$ (-- no $f_{\text{CE}}$, Table \ref{tab-edps-person-ablation}). This ablation mirrors the proposed approach of \citet[LACE]{mysore2023lace} for controllable recommendations. As expected, we see that the lack of query-document scores results in a large drop in performance indicating the value of \ctrlce over $f_{\text{Mem}}$ alone. 

\textbf{No mixing model.} In this experiment, we train a memory augmented cross-encoder without the mixing model $g_{\text{Mix}}$ (-- no $g_{\text{Mix}}$, Table \ref{tab-edps-person-ablation}) producing test time rankings using a simple summation of query-document and user-document scores: $s_d=s_d^q+s_d^u$. This may also be seen as a model resulting from stage-1 training alone. Here, we see that this approach consistently underperforms \ctrlceit and \ctrlcecv, indicating the value of $g_{\text{Mix}}$ for ranking performance.

\textbf{No calibrated objective.} Finally, we consider a model similar to \ctrlce, trained with $g_{\text{Mix}}$ with two-stage training but lacking in the calibrated softmax objective of Section \ref{sec-edps-memce-training} and instead using a standard softmax objective for both training stages (-- no calibration, Table \ref{tab-edps-person-ablation}). We see that omission of the calibrated objective results in a similar performance to \ctrlce showing calibrated training to not harm performance. In Section \ref{sec-edps-calibeval} we show how the calibrated training results in a stronger correlation between $w$ and the performance of $f_{\text{CE}}$ indicating its value for controllable personalization.

\begin{table}[t]
\centering 
\caption{\ctrlce compared to RetAugCE for the control action of ``no personalization''. MRR follows identical trends.}
\scalebox{0.85}{
\begin{tabular}{lcccc}
\toprule
 & \multicolumn{1}{c}{Comp Sci} & \multicolumn{1}{c}{Physics} & \multicolumn{1}{c}{Pol Sci} & \multicolumn{1}{c}{Psych}\\ 
\cmidrule(lr){2-2}\cmidrule(lr){3-3}\cmidrule(lr){4-4}\cmidrule(lr){5-5}
Model & \small{NDCG@10} & \small{NDCG@10} & \small{NDCG@10} & \small{NDCG@10}\\
\midrule
CrossEnc & 0.2934 & 0.3330 & 0.3102 & 0.3346\\
\midrule
RetAugCE & {0.3244} & {0.3691} & {0.3384} & {0.3778}\\
-- no personalization & 0.2264 & 0.2699 & 0.2039 & 0.2949\\
\midrule
\ctrlceit & {0.3223} & {0.3657} & {0.3378} & {0.3704}\\
-- no personalization & 0.3010 & 0.3440 & 0.3164 &0.3465\\
\ctrlcecv & {0.3118} & {0.3583} & {0.3310} & {0.3614}\\
-- no personalization & 0.2936 & 0.3344 & 0.3151 & 0.3368\\
\bottomrule
\end{tabular}
}
\label{tab-edps-person-ctrl}
\end{table}
\section{Interaction Evaluation}
\label{sec-edps-inteval}
\ctrlce supports control over personalized search in three ways: (1) support for a ``no personalization'' setting, (2) effectively highlighting queries where user edits to $\mathcal{P}_u$ may be necessary, (3) control over item or concept based profiles $\mathcal{P}_u$. 
Here, we demonstrate how \ctrlce effectively supports these control actions. In Section \ref{sec-edps-noperson}, we evaluate \ctrlce's ability to support a ``no personalization'' action compared against the RetAugCE model. In Section \ref{sec-edps-calibeval} we demonstrate that our mixing model closely tracks the performance of $f_{\text{CE}}$ allowing \ctrlce to obtain user edits only when necessary. Finally, in Section \ref{sec-edps-editeval} we demonstrate the controllability provided by editable item and concept-based profiles in a user study.

\subsection{``No personalization'' evaluation}
\label{sec-edps-noperson}
\subsubsection{Setup.} \ctrlce accomplishes ``no personalization'' by ranking documents using query-document score $s_d^q$ and dropping user-document score $s_d^u$. We compare this to the retrieval-augmented baseline RetAugCE. Here, ``no personalization'' is accomplished by only inputting query-document pairs into RetAugCE, dropping a retrieved document from $D_u$. In this setup, a controllable model must perform at par with a non-personalized cross-encoder.

\subsubsection{Results.} In Table \ref{tab-edps-person-ctrl} we see that dropping personalization from \ctrlce reverts it to perform similar to a non-personalized cross-encoder CrossEnc, indicating its ability to maintain performance while accomplishing a ``no personalization'' control action. On the other hand, RetAugCE sees a large drop in performance from not using the retrieved context indicating it to be a much harder model to control.  
Note also, that ranking for ``no personalization'' may be accomplished efficiently through per-query cached representations, not requiring repeated forward passes through \ctrlce.

\subsection{Selective control evaluation}
\label{sec-edps-calibeval}
\begin{figure}[t]
    \centering
    \subfloat[\centering \ctrlcecv in Comp Sci]{{\includegraphics[width=4cm]{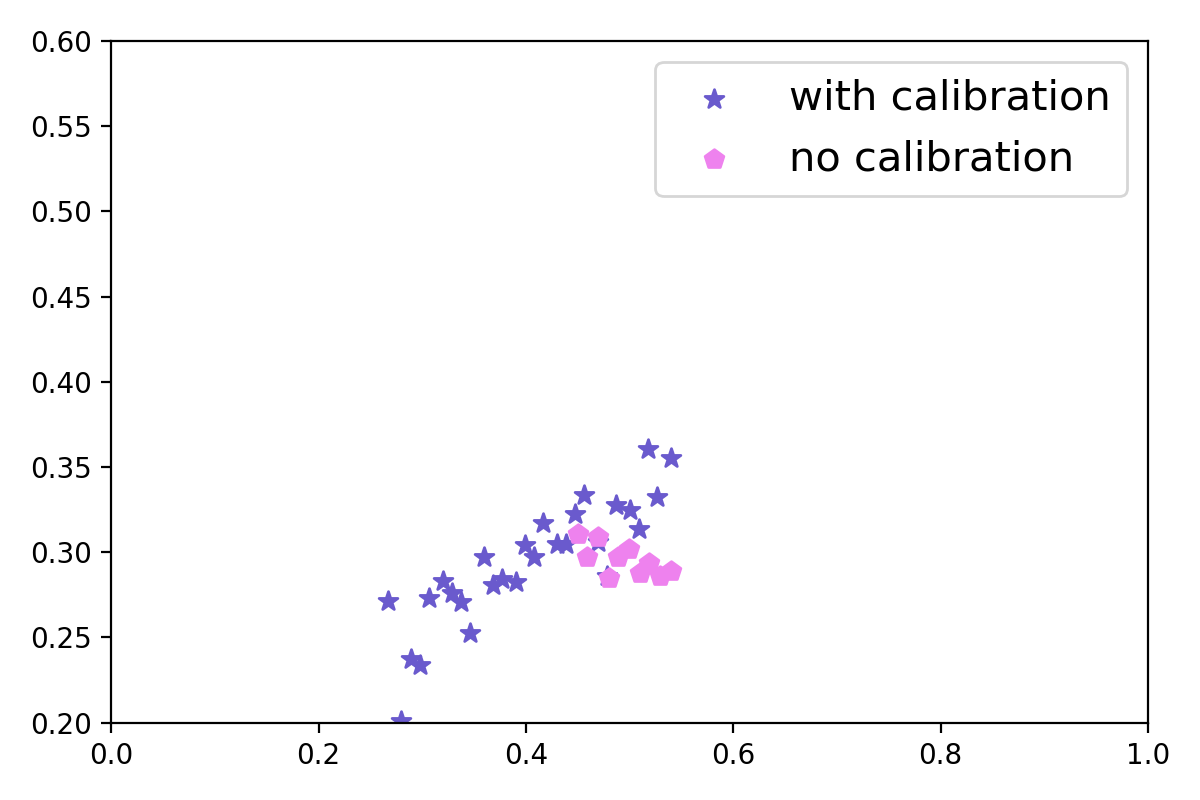}}}%
    ~
    \subfloat[\centering \ctrlcecv in Physics]{{\includegraphics[width=4cm]{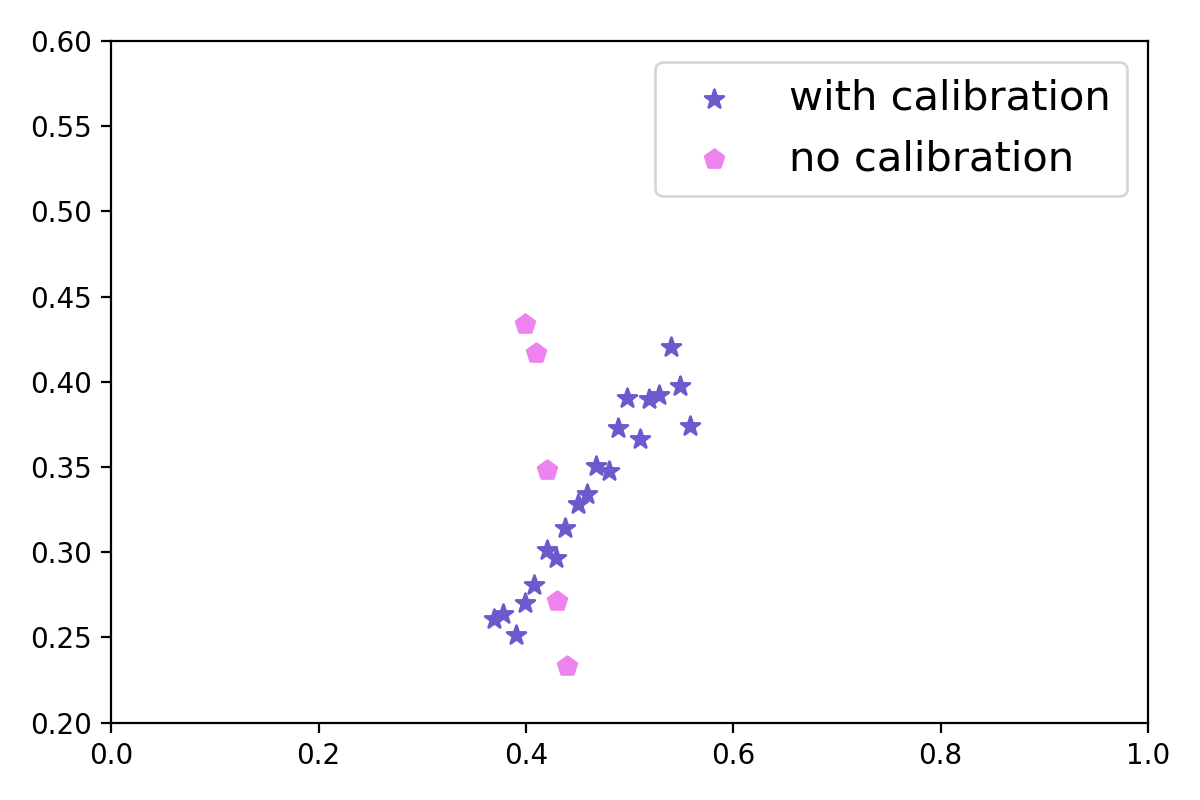}}}%
    
    \subfloat[\centering \ctrlcecv in Pol Sci]{{\includegraphics[width=4cm]{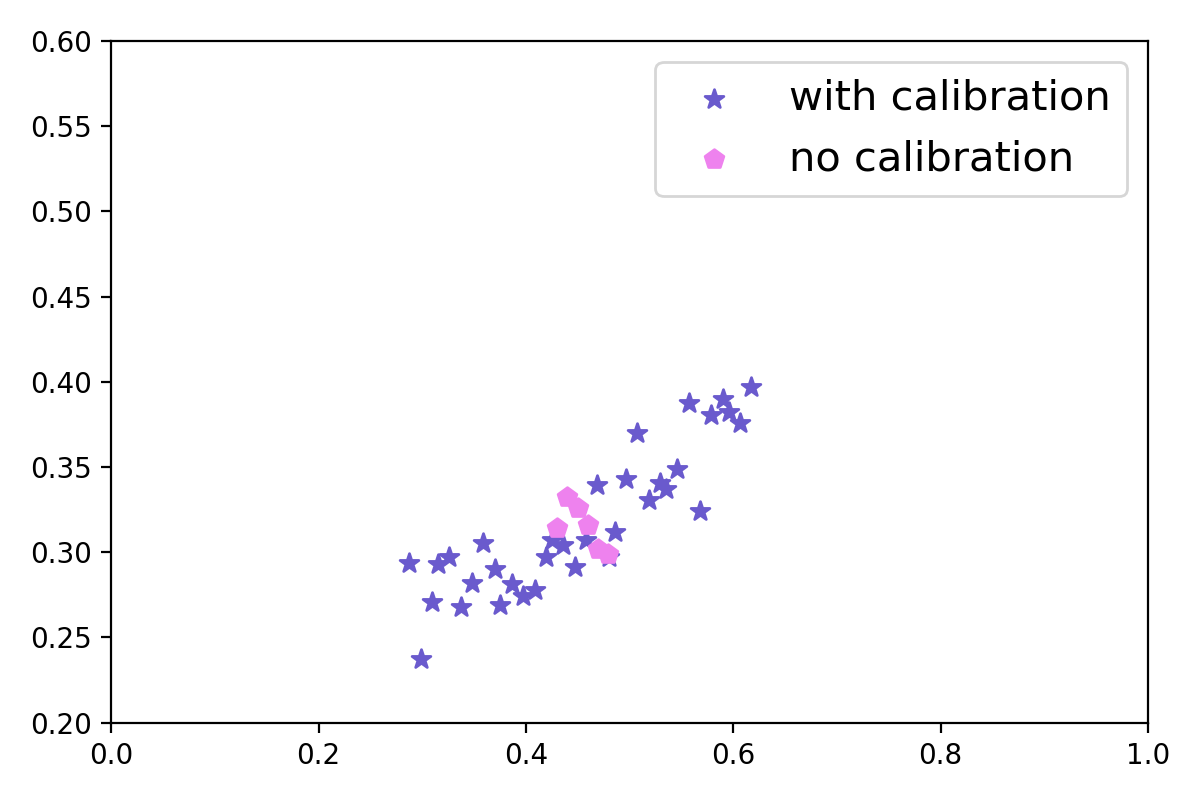}}}%
    ~
    \subfloat[\centering \ctrlcecv in Psych]{{\includegraphics[width=4cm]{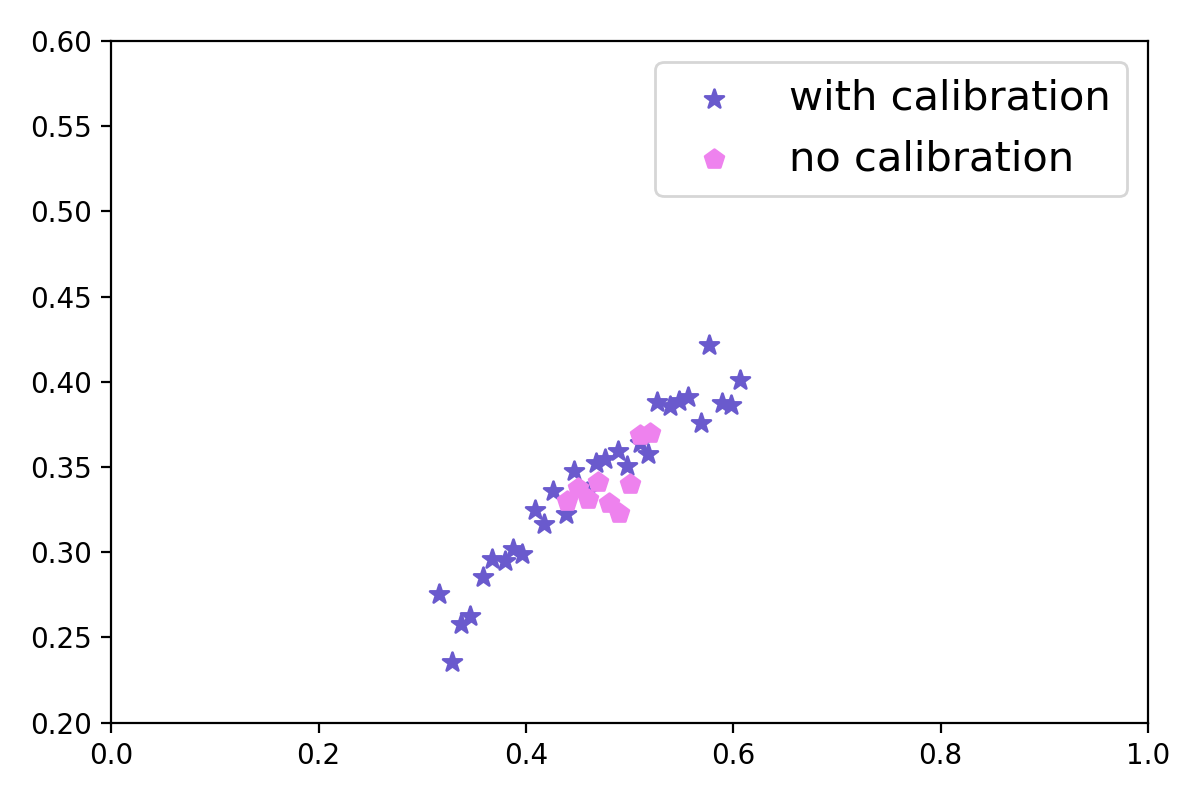}}}%
    \caption{Scores produced by the mixing model $g_{\text{Mix}}$ used to combine $f_{\text{CE}}$ and $f_{\text{Mem}}$ (Equation \eqref{eq-edps-highlevelscore}) plotted against the NDCG@10 for $f_{\text{CE}}$. \ctrlcecv (blue) is compared against a model trained without a calibrated objective (pink). Our calibrated objective ensures that $g_{\text{Mix}}$ scores are proportional to $f_{\text{CE}}$ performance. \ctrlceit shows identical trends.}
    \label{fig-edps-person-calibration-cv}%
\end{figure}

To demonstrate that \ctrlce selectively highlights the queries which would benefit from user control we show how the mixing model ($g_{\text{Mix}}$) also serves as a performance predictor for the cross-encoder model ($f_{\text{CE}}$) by evaluating its calibration performance -- i.e the ability of $g_{\text{Mix}}$ scores to be proportional to the ranking performance of $f_{\text{CE}}$. This enables system designers to use $g_{\text{Mix}}$ to identify queries where \ctrlce will rely more heavily on $f_{\text{Mem}}$ and where user over $\mathcal{P}_u$ may improve performance. We report results in Table \ref{tab-edps-person-calibration} and Figure \ref{fig-edps-person-calibration-cv}. Finally, we present a small-scale case study to highlight the queries which $g_{\text{Mix}}$ selects for personalization in Table \ref{fig-edps-person-calib-examples}.
\begin{table}
\centering 
\caption{The Pearson correlation between scores produced by the mixing model $g_{\text{Mix}}$ and NDCG@10 for $f_{\text{CE}}$. \ctrlceit and \ctrlcecv are compared against the respective models trained without a calibrating objective.}
\scalebox{0.85}{
\begin{tabular}{lcccc}
\toprule
 & \multicolumn{1}{c}{Comp Sci} & \multicolumn{1}{c}{Physics} & \multicolumn{1}{c}{Pol Sci} & \multicolumn{1}{c}{Psych}\\ 
\cmidrule(lr){2-2}\cmidrule(lr){3-3}\cmidrule(lr){4-4}\cmidrule(lr){5-5}
Model & $f_{\text{CE}}$ & $f_{\text{CE}}$ & $f_{\text{CE}}$ & $f_{\text{CE}}$\\
\toprule
\ctrlceit & \textbf{0.81} & \textbf{0.90} & \textbf{0.73} & \textbf{0.86}\\
-- no calibration & 0.13 & 0.54 & 0.48 & 0.62\\
\midrule
\ctrlcecv & \textbf{0.73} & \textbf{0.90} & \textbf{0.76} & \textbf{0.95}\\
-- no calibration & 0.22 & 0.92 & 0.42 & 0.38\\
\bottomrule
\end{tabular}
}
\label{tab-edps-person-calibration}
\end{table}

\subsubsection{Setup.} We measure calibration performance with the Pearson correlation between the score $w$ produced by $g_{\text{Mix}}$ for the top-1 document for each query and the NDCG@10 metric for $f_{\text{CE}}$. To compute this metric, we bucket all the queries in our test set into 100 equal-sized buckets based on the value of $w$ for the top retrieved document for a query. Then we compute the average NDCG@10 for the queries within each bucket. Finally, we measure the Pearson correlation between the lower edge of each bucket and the average NDCG@10 metric within each bucket while excluding buckets with fewer than 50 queries. We also plot these values in Figure \ref{fig-edps-person-calibration-cv}. We compare \ctrlce (blue) to models trained without the calibrated objective of Section \ref{sec-edps-memce-training} (pink). 
\begin{table}[]
\caption{Example queries which $g_{\text{Mix}}$ predicts as likely to require personalization ({red}). These queries have improved performance in \ctrlce over $f_{\text{CE}}$ (``Gain''). On the other hand, queries predicted as likely to perform well with $f_{\text{CE}}$ alone see no improvement from personalization ({green}).} 
\scalebox{0.75}{
\begin{tabular}{rlccl}
\toprule
 & & $g_{\text{Mix}}$ & Gain & Query\\
ID & Query text & score & \small{NDCG@10} & type \\
\midrule
\colorbox{pink}{CS1} & ``normal integration survey'' & {0.15} & {+0.41} &  {exploratory}\\
\colorbox{pink}{CS2} & ``tense data mining for data fusion'' & {0.08} & {+0.40} & {ambiguous}\\
\colorbox{green!25}{CS3} & \multicolumn{1}{p{5cm}}{``scale aware cnn pedestrian detection''} & {0.64} & {+0.00} & {unambiguous}\\
\colorbox{green!25}{CS4} & \multicolumn{1}{p{5cm}}{``katyusha acceleration sgd''} & {0.63} & {-0.01} & {unambiguous}\\
\colorbox{pink}{PS1} & \multicolumn{1}{p{5cm}}{``worldwide research on probiotics''} & {0.06} & {+0.25} & {exploratory}\\
\colorbox{pink}{PS2} & \multicolumn{1}{p{5cm}}{``consumers' willingness to pay for hale''} & {0.12} & {+0.22} & {ambiguous}\\
\colorbox{green!25}{PS3} & \multicolumn{1}{p{5cm}}{``patent pools, competition, and innovation, evidence from us industries''} & {0.50} & {+0.00} &  {unambiguous}\\
\colorbox{green!25}{PS4} & \multicolumn{1}{p{5cm}}{``co2 emissions in chinas lime industry''} & {0.45} & {-0.01} & {unambiguous}\\
\bottomrule
\end{tabular}}
\label{fig-edps-person-calib-examples}
\end{table}

\subsubsection{Results.} In Table \ref{tab-edps-person-calibration} we note that the calibrated training of \ctrlce results in $g_{\text{Mix}}$ being consistently linearly correlated with the performance of $f_\text{CE}$. We also note from Figure \ref{fig-edps-person-calibration-cv} that in the absence of calibrated training the $g_{\text{Mix}}$ scores (pink) rarely track the performance of $f_{CE}$. This not only results in poor performance (see Table \ref{tab-edps-person-ablation}, ``-- no calibration'' row) but also indicates that the scores would not be useful for selectively soliciting user edits. The strong correlation of the $g_{\text{Mix}}$ with $f_{\text{CE}}$ indicates its potential for guiding users to provide control interactions for $\mathcal{P}_u$ only when $f_{\text{CE}}$ alone is likely to underperform.

\subsubsection{Case study.} In Table \ref{fig-edps-person-calib-examples} we include a small-scale case study of $g_{\text{Mix}}$ outputs to illustrate queries selected for personalization. Here we manually examine the queries that receive among the highest and lowest weights from $g_{\text{Mix}}$. For these queries, we examined the ranked documents by $f_{\text{CE}}$ and \ctrlce (i.e $w\cdot f_{\text{CE}} + (1-w)\cdot f_{\text{Mem}}$) and their relevance judgments. Queries with a low  $g_{\text{Mix}}$ predicted weight are likely to require personalization and in turn benefit from user edits to $\mathcal{P}_u$. We show examples from \ctrlceit predictions for computer science and political science domains given the relative ease of understanding these domains.

In Table \ref{fig-edps-person-calib-examples} we see that $g_{\text{Mix}}$ commonly scores exploratory queries (\colorbox{pink}{CS1}, \colorbox{pink}{PS1}) and ambiguous look-up queries (\colorbox{pink}{CS2}, \colorbox{pink}{PS2}) with a low score. In both these cases personalization based on $f_{\text{Mem}}$ plays a greater role in producing ranked documents. In both these cases we see that $f_{\text{Mem}}$ helps \ctrlce outperform $f_{\text{CE}}$ alone (positive ``CCE gain''). We also see that unambiguous look-up queries that seek more specific relevant documents (\colorbox{green!25}{CS3-4}, \colorbox{green!25}{PS3-4}) result in higher scores from $g_{\text{Mix}}$ for $f_{\text{CE}}$. Consequently $f_{\text{CE}}$ and \ctrlce perform nearly identically in these queries. This illustrates that $g_{\text{Mix}}$ successfully identifies queries that benefit from personalization and could benefit from user edits to $\mathcal{P}_u$. Finally, we note that not all queries with low values of $w$ may benefit from edits to $\mathcal{P}_u$, e.g.\ some could benefit from query reformulation. We leave the exploration of finer-grained performance prediction to future work. Nevertheless, our experiments show that $g_{\text{Mix}}$ remains calibrated and enables obtaining user control for $\mathcal{P}_u$ only when necessary.

\subsection{Editability evaluation}
\label{sec-edps-editeval}
We evaluate the controllability of the editable memories of \ctrlce in a user study with realistic queries and user profiles. We ran our user study with expert annotators interacting with interactive prototypes of \ctrlceit and \ctrlcecv. Through our study, we aim to answer the research question: Are users able to improve the search performance for \ctrlceit and \ctrlcecv through interaction with user profiles, $\mathcal{P}_u$? Table \ref{tab-ustudy-ranking} and Section \ref{sec-edps-ustudy-results} present our results.

\subsubsection{Data and Model Setup.} 
Our user study was run as an expert-driven user study with two computer science annotators. Our annotators interacted with an interactive prototype of \ctrlce and judged 45 realistic queries and their associated user profiles selected from our Comp Sci evaluation dataset.  
Because we are primarily interested in evaluating the controllability of editable memories in \ctrlce we selected the 200 queries where $g_{\text{Mix}}$ scores indicated that the query was most likely to benefit from personalization. This set of queries is also likely to benefit from user control to $\mathcal{P}_u$. Next, to prevent burdensome annotations we only retained queries that had between 20-50 historical documents ($D_u$). Next, for the 200 queries, our expert annotators indicated which queries were topically familiar and of research interest to them -- allowing them to reasonably stand in as the original users for judging \ctrlce rankings. This resulted in 45 queries for our user study. Our expert annotators were computer science researchers and part of the authorship team. Both annotators had interests in design and AI, had experience reading research papers, and were compensated for their time. We opted for an expert-driven user study instead of one with real users due to limits on the length of our user study. 
Because not all queries require personalization a study that relies on real users would need to be run for a longer period to ensure that users organically made sufficient queries which required personalization and edits to their user profile. Instead, we conduct careful evaluations of all parts of \ctrlce (Section \ref{sec-edps-noperson}-\ref{sec-edps-studyprocedure}) and leave end-to-end evaluations over longer times to future work.
\begin{table}
\centering 
\caption{User study evaluation of \ctrlceit and \ctrlcecv demonstrating their ability to enable users to improve search performance through interactions with a user profile.}
\scalebox{1}{
\begin{tabular}{lcccc}
\toprule
        & \multicolumn{2}{c}{\ctrlceit} & \multicolumn{2}{c}{\ctrlcecv} \\
        \cmidrule(lr){2-3}\cmidrule(lr){4-5}
        & \small{NDCG@20} & \small{R@20} & \small{NDCG@20} & \small{R@20}\\
\toprule
Initial & 0.7169 & 0.7549 & 0.7169 & 0.7549\\
Tuned   & \textbf{0.7537} & \textbf{0.8247}$^{*}$ & \textbf{0.7303} & \textbf{0.8139}$^{**}$\\
Gain    & +0.0368 & +0.0698 & +0.0134 & +0.0590\\
\bottomrule
\end{tabular}
}
\label{tab-ustudy-ranking}
\end{table}

\subsubsection{Study Procedure.}
\label{sec-edps-studyprocedure}
To evaluate the controllability of the \ctrlce models our user study was conducted in three phases: (A) Our annotators rated an initial ranked list from \ctrlceit and \ctrlcecv for the relevance of documents, (B) Next, they tuned user profiles with interactions (Section \ref{sec-edps-memce-inference}) which they judged would improve the initial ranked list, and (C) They rated an updated ranked list of documents produced as a result of the control action. The ratings gathered from this procedure were used to measure the effect of \ctrlce's editable memories on search performance. To ensure that our annotators made reasonable and unbiased ratings, we included a guideline generation and agreement measurement phase in our study. Further, we release the guidelines and generated ratings in our code release to ensure transparency in the process. To measure agreement, both annotators rated 15 shared queries and the ranked lists from \ctrlceit. Then, they met and resolved their rating disagreements, and created an adjudicated set of ratings and rating guidelines. We noted annotator agreement with the adjudicated ratings to be Cohens $\kappa = 0.81$ and a rank-correlation of $\rho = 0.84$, indicating the annotation guideline to be sound and the annotators in agreement. The guideline was then applied to 30 queries not used for agreement measurements, and Table \ref{tab-ustudy-ranking} reports these. In making their ratings annotators first familiarized themselves with the user profile and query, and then rated $K=30$ ranked documents on a 3-point scale. They reported spending 15-20 minutes per query.

\subsubsection{Results.} Table \ref{tab-ustudy-ranking} presents the primary result of our user study. We report NDCG and Recall at deeper ranks, $K=20$ to illustrate how control over personalization improves the users' ability to explore collections -- this is commonly done through exploring to deeper rank positions. Because we gather ratings for $K=30$ we report Recall@20. We report statistical significance at $p < 0.05$ and $p < 0.10$ with a paired t-test, denoted as $^{*}$ and $^{**}$. 

From Table \ref{tab-ustudy-ranking} we note that users were able to improve the performance of \ctrlceit by 5-10\% and \ctrlcecv by 2-8\% across metrics, with statistically significant improvements for R@20. Based on this we answer our research question in the affirmative - users are able to effectively interact with editable memories in \ctrlce to improve search performance. Further, the larger improvements in R@20 indicate that control interactions improve exploratory ability than precision-oriented performance. We also note that larger-scale studies may be needed to establish statistically significant improvements in NDCG@20. Finally, our user study did not probe aspects of interface usability or user trust from controllable personalization, future work may probe these aspects further in longer-term deployments in realistic application contexts.
\label{sec-edps-ustudy-results}

\section{Conclusion}
In this paper, we introduce \ctrlce, a memory-augmented cross-encoder for controllable personalized search. To facilitate control while achieving strong ranking performance, we augment expressive cross-encoder models with editable memories of user documents. To ensure that our expressive cross-encoder is able to interact with a multi-vector user memory we formulate it as a novel embedding cross-encoder. Further, we introduce a calibrated mixing model that indicates when queries benefit from personalization and in turn from greater user interactions. In experiments on four scientific domains, we demonstrate \ctrlce to improve upon a wide variety of standard prior methods spanning, sparse, dense, cross-encoder, and personalized approaches. We demonstrate the controllability of \ctrlce through experiments demonstrating its ability to support a ``no personalization'' interaction. In calibration evaluations and a case study, we demonstrate its ability to seek user interaction only when necessary. Finally, in a user study we demonstrate that when user interactions are sought, interaction with item and concept-based profiles successfully improves performance.

\begin{acks}
We thank anonymous reviewers for their feedback. This work was partly supported by the Center for Intelligent Information Retrieval, IBM Research AI through the AI Horizons Network, CZI under the project Scientific Knowledge Base Construction, NSF grant number IIS-2106391, Amazon Alexa Prize Competition, Lowes, Adobe, Google, and Microsoft. Any opinions, findings, conclusions, or recommendations expressed in this material are those of the authors and do not necessarily reflect those of the sponsors.
\end{acks}

\bibliographystyle{ACM-Reference-Format}
\balance
\bibliography{ed_exp-short}

\end{document}